\documentclass{svproc}
\usepackage{graphicx}
\usepackage{cancel}
\usepackage{color}

\newcommand{\tcb}{\textcolor{blue}}
\usepackage{url}

\usepackage{amsmath}
\usepackage{amssymb}
\usepackage{mathrsfs}
\usepackage{ae,aecompl}
\usepackage{color}
\usepackage{fancyhdr}
\usepackage{float}
\floatstyle{ruled}
\newfloat{algorithm}{hbt}{alg}
\floatname{algorithm}{Algorithm}
\floatstyle{boxed}

\usepackage{cite}
\usepackage{cancel}
\usepackage[english]{babel}
\usepackage{multirow}
\begin{document}
	\mainmatter              % start of a contribution
	\title{Identification of melanoma diseases from multispectral dermatological images using a novel BSS approach}
	\titlerunning{Identification of melanoma diseases from multispectral dermatological images using a novel BSS approach}  % abbreviated title (for running head)
	\author{Mustapha Zokay \and Hicham Saylani}
	\tocauthor{Hicham Saylani}
	\institute{Laboratoire de matériaux, signaux et systèmes et modélisation physique,\\ Faculté des Sciences, Université Ibn Zohr, BP 8106, Cité Dakhla, Agadir, Morocco
		\email{mustapha.zokay@edu.uiz.ac.ma,} \hspace{0.1cm}
		\email{h.saylani@uiz.ac.ma}}

	\maketitle              % typeset the title of the contribution

	%===========================================================
	\begin{abstract}
		In this paper\footnote{The paper has been accepted at the \textbf{MICAD 2022} conference and is currently being published as a chapter in the Springer book : \textbf{ Lecture Notes in Electrical Engineering.}} we propose a new approach to identify melanoma diseases by identifying the distribution of its main skin chromophores (melanin, oxyhemoglobin and deoxyhemoglobin) from multispectral dermatological images. Based on Blind Source Separation (BSS), our approach takes into account the shading present in most of the images. Assuming that the multispectral images have at least 4 spectral bands, it allows to estimate the distribution of each chromophore in addition to the shading without any a priori information, contrary to all existing methods that use 3 bands, i.e. RGB images. Indeed, the fact of neglecting the shading degrades their performance. To validate our method, we used a database of real multispectral dermatological images of skin affected by \textit{melanoma} cancer. To measure our performance, in addition to the classical criterion of visually analyzing the estimated distributions with referring to the physiological knowledge of the disease, we proposed a new criterion that is based on our independence hypothesis. Using these two criteria, we could see that our approach is very efficient for the identification of melanoma.
	\end{abstract}
	\section{Introduction}
	The skin is the largest organ in the human body. It contains three main chromophores which are melanin, oxyhemoglobin and deoxyhemoglobin. The distribution of these chromophores is of great importance for dermatologists who use it for the identification and early detection of skin diseases. One of the techniques that is increasingly used and has proven its effectiveness in identifying the distribution of chromophores is multispectral imaging \cite{jacques2010rapid}.
	However, the direct use of multispectral images tends to give erroneous information on the distribution of chromophores.
	Indeed, the light intensity reflected by the skin does not only depend on these three chromophores but also on the geometry of the skin surface, called shading.
	The multispectral images obtained at different wavelengths can therefore be considered as mixtures of four constituents which are the three chromophores and the shading.
	Thus, if we consider these constituents as sources, this problem can be seen as a source separation problem, known as an inverse problem that belongs to the field of signal processing.
	The objective of source separation is to estimate the sources from the sole knowledge of their mixtures. As this estimation is done without any a priori information neither on the sources nor on the mixing coefficients, it is called Blind Source Separation (BSS). 
	It is easy to see that the BSS problem is an ill-posed inverse problem which admits an infinity of solutions so it is essential to add hypotheses on the sources and/or on the mixing coefficients, which gave rise to 3 main families of BSS methods which are the \textit{Independent Component Analysis} (ICA), the \textit{Sparse Component Analysis} (SCA) and the \textit{Non-negative Matrix Factorization} (NMF) (see \cite{book} for more details). 
	During the last decade, the use of BSS methods for non-invasive identification of chromophore distributions has attracted the interest of several researchers \cite{jolivot2012quantification,liu2013melanin,ojima2004application,spigulis2015snapshot,spigulis2017smartphone,kuzmina2010multispectral,gong2012hemoglobin} who all adopted the same mixing model.
	Knowing that most of these researchers were interested in RGB images and thus in 3 wavelengths $\lambda_{i}$, $i\in$\{1,2,3\} which represent respectively the central wavelengths of the Blue, Green and Red bands, and if we note $j\in$\{$1,2,3$\} the index related to the chromophores and which represents respectively  
	melanin, oxyhemoglobin and deoxyhemoglobin, then the classical mixing model is written:
	\begin{equation}\label{eq0}
		I_{\lambda_{i}}(\textit{\textbf{u}})=\sum_{j=1}^{j=3}a_{ij}.S_j(\textit{\textbf{u}})+p_{d}(\textit{\textbf{u}})+n_i,  \hspace*{0.4cm} i\in\{1,2,3\},
		\vspace*{-0.5cm}
	\end{equation}
	where: 
	\renewcommand{\labelitemi}{$\bullet$}
	\begin{itemize}
		\item $I_{\lambda_{i}}(\textit{\textbf{u}})$ is the logarithm inverse of the reflectance detected by the camera, at wavelength $\lambda_i$, at the pixel of coordinates $(x,y)=\textit{\textbf{u}}$,
		\item 	$S_j(\textit{\textbf{u}})$ represents the chromophore of index $j$,  
		\item $a_{ij}$ represents the mixing coefficient which depends on the molar absorption coefficient of the chromophore $S_j(\textit{\textbf{u}})$ and the light penetration depth in the skin at wavelength $\lambda_i$,
		\item $p_{d}(\textit{\textbf{u}})$ represents the shading variation in the image,
		\item $n_{i}$ represents the characteristics of the sensor.
	\end{itemize}
	
	This mixing model has been adopted by all existing BSS methods \cite{jolivot2012quantification,liu2013melanin,ojima2004application,spigulis2015snapshot,spigulis2017smartphone,kuzmina2010multispectral}, but they differ in the assumptions made about the chromophores and the procedure followed to reach the final objective, so that these methods can be classified into two main families. 
	The first family includes the BSS methods that consider oxyhemoglobin and deoxyhemoglobin as a single source called hemoglobin \cite{jolivot2012quantification,liu2013melanin,ojima2004application, mitra2010source,gong2012hemoglobin}.
	Indeed, in \cite{ojima2004application, mitra2010source}, the authors applied \textit{Principal Component Analysis} (\textit{PCA}) and then \textit{ICA} on an RGB image of the skin, assuming that the shading is constant throughout the image and that melanin and hemoglobin are independent.
	In \cite{inbook}, Madooei et al. proposed a new 2-D color chromaticity to eliminate shading using the geometric mean color, and then estimated the distributions of the two chomophores using \textit{ICA}. 
	In \cite{gong2012hemoglobin}, Gong et al. proposed to estimate the distributions of both chromophores from an RGB image using \textit{NMF}.
	In \cite{jolivot2012quantification}, Galeano et al. proposed to use a neural network-based system and then applied \textit{NMF} to separate the melanin from the hemoglobin, but in their model they neglected the shading.\\
	The second family includes the BSS methods which are based on a priori information on the absorption spectra of the three chromophores and the light penetration depth in the skin \cite{spigulis2015snapshot,spigulis2017smartphone}.
	the authors proposed to remove the shading and the specular reflection from the RGB image using respectively white paper and polarizers, then they relied on knowledge of absorption coefficients and estimates the light penetration depth in order to deduce an empirical mixing matrix to extract the distributions of the three chromophores. It should be noted, that the weakness of this family lies in the level of accurate estimation of light penetration depth into the skin.
	
	In this paper, we propose a new method based on BSS to estimate the distribution of all the  three main chromophores separately, in addition to the shading distribution which we consider as a full-fledged source, unlike all existing methods. Based on more realistic assumptions and applying to multispectral images with at least 4 spectral bands, our new method exploits the intrinsic properties  
	of each chromophore.
	To validate our method we use a database of real multispectral dermatological images of skin affected by melanoma cancer disease \cite{cin-Lezoray-2014-1}.
	For the performance measurement, in addition to the classical criterion which is based on the visual analysis of the estimated distributions of the three chromophores,  we propose in this paper a new numerical criterion which is based on the measure of independence between the estimated distributions of melanin and hemoglobin.
	The rest of this paper is organized as follows. Section 2 presents our new method for estimating the distribution of the three chromophores and shading. Section 3 presents the results of the tests performed  followed by a last section dedicated to a conclusion and perspectives for our work.
	\section{Proposed method}
	The first key idea of our new method is to consider shading as a full-fledged source, in addition to the three sources of interest, which allows us to avoid the unrealistic assumption made by most existing methods that its contribution is the same at all wavelengths\footnote{Indeed, we found from the experimental curve obtained by \textit{PCA}
		used in \cite{eguizabal2013direct} that the contribution of shading ($p_{d}$) is not equal to $1$ in all mixtures.}.
	However, as the number of sources involved becomes equal to 4 we are interested in this paper for the case where we have multispectral images with at least 4 bands, said case determined\footnote{This is the case where we have as many mixtures as sources. On the other hand, in the case where we have more mixtures than sources, called the over-determined case, we can easily return to the determined case by applying a \textit{PCA}.}.
	From equation ({\ref{eq0}}), we can see that the term $n_{i}$  does not give any information on chromophore distributions. Thus, as in \cite{ojima2004application}, we begin by eliminating it from our mixtures based on the following hypothesis \textbf{(H1)}.
	\begin{description}
		\item[(H1) :] There is at least one pixel in the 4 images where the concentrations of the three chromophores and shading are all zero, i.e.:
		\begin{equation}\label{EqH1}
			\exists\,\, \textit{\textbf{u}} \, \,/ \,\,n_{i}=min(I_{\lambda_{i}}(\textit{\textbf{u}})).
		\end{equation}
	\end{description}
	So, the new mixture model is written:
	\begin{equation}\label{7}
		X_{i}(\textit{\textbf{u}})=\sum_{j=1}^{j=3}a_{ij} \cdot S_j(\textit{\textbf{u}})+a_{i4}\cdot S_{4}(\textit{\textbf{u}}),  \hspace*{0.4cm} i\in[1,4] %\{1,2,3,4\},
		\vspace*{-0.3cm}
	\end{equation}
	where $\,\,X_{i}(\textit{\textbf{u}})=I_{\lambda_{i}}(\textit{\textbf{u}})-n_{i}$,
	$S_{4}(\textit{\textbf{u}})=p_{d}(\textit{\textbf{u}})$ and
	$a_{i4}\in \mathbb{R}$.
	
	\vspace*{0.1cm}
	As for all BSS methods, we generate new 1D mixtures (vectors), which we note $X_{i}(v)$, from the 2D mixtures (images) $X_{i}(\textit{\textbf{u}})$ by concatenating the rows of the latter. We then have:
		\vspace*{-0.2cm}
	\begin{equation}
		X_{i}(v)=\text{vec}(X_{i}(\textit{\textbf{u}}))=\sum_{j=1}^{j=4}a_{ij} \cdot S_j(v), \hspace*{0.4cm} i\in[1,4] 
	\end{equation}
	
The second key idea of our method is to treat mixtures in two steps, unlike all existing methods that treat all mixtures at the same time. Indeed, we start by treating only two mixtures that contain only melanin and shading in order to separate them first, and then we eliminate their contribution from the other two mixtures remaining to keep only oxyhemoglobin and deoxyhemoglobin. These last two chromophores are then separated in a last step.
	These three steps of our method are detailed below.
	\subsection*{Step 1: Separation of sources $S_1(v)$ and $S_4(v)$}
	In this step we exploit the properties of each chromophore concerning spectral absorption as a function of the wavelength. Indeed, based on data published in \cite{van1989skin}, 
	we found that light absorption at wavelengths greater than $620\, nm$ is dominated by melanin, so the absorption coefficients $a_{ij}$
	of oxyhemoglobin and deoxyhemoglobin are all negligible, i.e. we have $a_{32}=a_{33}=0$ and $a_{42}=a_{43}=0$.
	Thus, the mixtures corresponding to the red and infrared bands can be re-written as follows:
	\begin{equation}\label{eq1}
		\left\lbrace
		\begin{array}{ll}
			X_{3}(v)=a_{31}\cdot S_1(v)+a_{34}\cdot S_4(v)\\
			X_{4}(v)=a_{41}\cdot S_1(v)+a_{44}\cdot S_4(v)
		\end{array}
		\right.	
	\end{equation}

	We can re-write the equation system (\ref{eq1}) in a matrix form as follows:
	\begin{equation}
		{\bf X}(v)={\bf A}\cdot {\bf S}(v),
	\end{equation} 
where
	${\bf X}(v)=[X_{3}(v),X_{4}(v)]^{T}$, ${\bf S}(v)=[S_{1}(v),S_{4}(v)]^{T}$
	and ${\bf A}=\begin{pmatrix} 
		a_{31} & a_{34}\\
		a_{41} & a_{44}\\
	\end{pmatrix}$.
	
	In this first step we assumed that the two sources $S_{1}(v)$ and  $S_4(v)$ are independent, in which case we can use one of the \textit{ICA} methods to separate them. We have opted here for the \textit{AMUSE} method \cite{tong1991indeterminacy} for its simplicity since it exploits only the second order statistics of the signals. 
	%%%%%%%%%%%%%%%%%%
	Indeed, the working hypotheses of this method are the following.
	%%%%%%%%%%%%%%%%%%
	\begin{description}
		\item[(H2) :] The sources $S
		_{j}(v)$ are {\it auto-correlated} and {\it mutually uncorrelated}, i.e. :
		\begin{equation}\label{EqH4}
		\forall \,\, \tau,\, \hspace{0.3cm} \left\lbrace 
		\begin{array}{ll}
		\hspace{0.1cm} E[S_j(v)\cdot S_j(v-\tau)] \neq 0, \,\,\,\,	j\in \{1,4\}\\
		\hspace{0.1cm} E[S_1(v)\cdot S_4(v-\tau)]=E[S_1(v)]\cdot E[S_4(v-\tau)] \,\,\,	
		\end{array}
		\right.	
		\end{equation}
		\item[(H3) :]  The \textit{condition of identifiability} for the method is verified, i.e. :
		\begin{equation}\label{EqH5}
		\exists \,\, \tau\neq 0 \, \,/\hspace{0.3cm} 
		\frac{E[S_1(v)\cdot S_1(v-\tau)]}{E[S_1^2(v)]} \neq
		\frac{E[S_4(v)\cdot S_4(v-\tau)]}{E[S_4^2(v)]}.
		\end{equation}
	\end{description}

The method \textit{AMUSE} allows us to estimate the separation matrix ${\bf A}^{-1}$ to a permutation matrix ${\bf P}$ and a diagonal matrix ${\bf D}$ \cite{tong1991indeterminacy}. By noting this matrix ${\bf C}= {\bf P}{\bf D}{\bf A}^{-1}$, we obtain finally 
the source matrix $ {\bf S}(v)$ with the same indeterminations as follows:		
	\begin{equation}
		{\bf C}\cdot {\bf X}(v)=\left( {\bf PDA}^{-1}\right) \cdot \left( {\bf AS}(v)\right) ={\bf PD}{\bf S}(v).
		%= {\bf Y}(v)
	\end{equation}
	
	Indeed, by noting ${\bf PD}{\bf S}(v)={\bf Y}(v)=[Y_{1}(v),Y_{4}(v)]$ and omitting the permutation\footnote{Indeed, the permutation matrix $\bf P$ can be identified based on the visual analysis 
			since the shading source $S_4(v)$ is easily differentiable.} 
	we have:
	\begin{equation}\label{EqYj0}
		Y_{j}(v)=\alpha_{j}\cdot S_{j}(v), \hspace*{0.2cm}j=1,4,
	\end{equation}
where the $\alpha_{j}$ are the elements constituting the diagonal of the matrix ${\bf D}$.

\subsection*{Step 2: Removal of $S_1(v)$ and $S_4(v)$ sources from mixtures}
	The goal of this step is to eliminate the contributions of the sources estimated $S_1(v)$ and $S_4(v)$ from the mixtures $X_{1}(v)$ and $X_{2}(v)$. For this we exploit the following independence hypothesis.
	\begin{description}
		\item[(H4) :] $S_{i}(v)$ and $S_{j}(v)$ are {\it mutually uncorrelated instantaneously}, for $i\in\left\lbrace 2,3\right\rbrace $
		and 
		$j\in\left\lbrace 1,4\right\rbrace$,
		i.e.: 
		\begin{equation}\label{EqH3}
			E[\tilde{S}_i(v)\cdot \tilde{S}_j(v)]=	E[\tilde{S}_i(v)]\cdot	E[\tilde{S}_j(v)]=0, \hspace*{0.1cm} \forall (i,j)\in\left\lbrace 2,3\right\rbrace  \times \left\lbrace 1,4\right\rbrace 
		\end{equation}
	\end{description}
where $\tilde{S}_i(v)$ are the centered versions\footnote{i.e. : $\tilde{S}_i(v)=S_i(v)-E[S_i(v)]$.}
of the sources $S_i(v)$. By denoting respectively $\tilde{X_{i}}(v)$ and $\tilde{Y_{j}}(v)$ the centered versions of the signals $X_{i}(v)$ and $Y_{j}(v)$, and by exploiting the relations (\ref{EqYj0}) and (\ref{EqH3}) we can write:
	\begin{eqnarray}
	E[\tilde{X_{i}}(v)\cdot \tilde{Y}_{j}(v)]&=& E\left[ \left( \sum_{k=1}^{k=4}a_{ik} \cdot \tilde{S}_k(v)\right) \cdot \left( \alpha_j\cdot \tilde{S}_{j}(v)\right) \right] \\
	&=&a_{ij} \cdot\alpha_j \cdot E[  \tilde{S}_j^2(v)] \label{EqXiYj}
	\end{eqnarray}
	
	On the other hand, according to the relation (\ref{EqYj0}) we have:
	\begin{equation}\label{EqYj2}
		E[\tilde{Y}_{j}^{2}(v)]=\alpha_{j}^{2} \cdot E[\tilde{S_{j}^{2}}(v)], \hspace*{0.4cm}j=1,4.
	\end{equation}
	
	Thus, by exploiting the relations (\ref{EqXiYj}) and (\ref{EqYj2}) we 
	can generate two new mixtures $Z_{i}(v) (i=1,2)$ which contain only the sources $S_2(v)$ and $S_3(v)$ as follows:
	\vspace*{-0.3cm}
	\begin{eqnarray}
		Z_{i}(v)&=&X_{i}(v)-\frac{E[\tilde{X_{i}}(v)\cdot \tilde{Y_{1}}(v)]}{E[\tilde{Y_{1}^{2}}(v)]}\cdot Y_{1}(v)-\frac{E[\tilde{X_{i}}(v)\cdot \tilde{Y_{4}}(v)]}{E[\tilde{Y_{4}^{2}}(v)]}\cdot Y_{4}(v) \\
	&=&X_{i}(v)-\frac{a_{i1}}{\alpha_{1}}\cdot (\alpha_1.\cdot S_1(v))-\frac{a_{i4}}{\alpha_{4}}\cdot (\alpha_4\cdot S_4(v))\\
		&=&a_{i2}\cdot S_{2}(v)+a_{i3}\cdot S_{3}(v) \label{EqZi}
	\end{eqnarray}
	\subsection*{Step 3 : Separation of Sources $S_2(v)$ and $S_3(v)$}
	The goal of this step is to separate the remaining $S_2(v)$ and $S_3(v)$ sources that represent oxyhemoglobin and deoxyhemoglobin respectively, and this time by treating the new mixtures $ Z_ {1} (v) $ and $ Z_ {2} (v) $ provided by the previous step.
	As these two sources are dependent, which is why most of the existing methods fail to separate them, we have opted in this paper for a new solution based on the exploitation of their sparsity\footnote{A source is said to be sparse in a given representation domain if some of its samples are zero in this domain.}.
	Here is our working hypothesis for this step.
	\begin{description}
		\item[(H5) :] For each source, there is at least one spatial area over which it is active while the other source is inactive, i.e.:		
		\begin{equation}
		\left\lbrace
		\begin{array}{ll}
			\exists \,V_1 \,/\, 
	\forall \,v\in V_1, \,\,{ S_{2}(v)=0} \,\,\,\,\text{and}\,\,\,\, {S_{3}(v)\ne 0}\\
				\exists \, V_2 \,/\,
	\forall \, v\in V_2, \,\,{S_{3}(v)=0} \,\,\,\,\text{and}\,\, \,\,{S_{2}(v)\ne 0}\label{EqSCA}
		\end{array}
		\right.	
		\end{equation}	
	\end{description}
	
	There are several BSS methods that exploit this sparsity assumption to achieve separation. We opted here for the \textit{TEMPROM} method proposed in\cite{abrard2005time} for its simplicity.
	This method consists of identifying the single-source areas $V_1$ and $V_2$ at first, and then calculating the ratio between the mixtures $Z_2(v)$ and $Z_1(v)$ on these areas at a second time, which ultimately allows to estimate the matrix of mixture involved. Indeed, by exploiting the two equations of the system (\ref{EqSCA}) and the relation (\ref{EqZi}) we obtain:
	%%%%%%%%%%%%%%%%%%%%%%%%%%%%%%%%%%
	\begin{equation}\label{eqr_i}
	\left\lbrace
	\begin{array}{ll}
	\forall \, v\in V_1, \,\,\, \frac{Z_{2}(v)}{Z_{1}(v)}=\frac{a_{23}\cdot S_{3}(v)}{a_{13} \cdot	S_{3}(v)}=\frac{a_{23}}{a_{13}}=r_{1}\\
		\forall \, v\in V_2, \,\,\, \frac{Z_{2}(v)}{Z_{1}(v)}=\frac{a_{22}\cdot S_{2}(v)}{a_{12}\cdot S_{2}(v)}=\frac{a_{22}}{a_{12}}=r_{2}
	\end{array}
	\right.	
	\end{equation}
	%%%%%%%%%%%%%%%%%%%%%%%%%
	Finally, by exploiting the relations (\ref{EqZi})
	%(\ref{Eqr1}) and (\ref{Eqr2}) 
	and \eqref{eqr_i} we obtain:
	%%%%%%%%%%%%%%%%%%%%%%%%%%%%%%%%
			\begin{equation} 
	\left\lbrace
	\begin{array}{ll}
		r_{1}\cdot Z_{1}(v)-Z_{2}(v)=(r_{1}\cdot a_{12}-a_{22})\cdot S_{2}(v)=\alpha_{2}\cdot S_{2}(v)=Y_{2}(v)\\
		r_{2}\cdot Z_{1}(v)-Z_{2}(v)=(r_{2}\cdot a_{13}-a_{23})\cdot S_{3}(v)=\alpha_{3}\cdot S_{3}(v)=Y_{3}(v)
	\end{array}
	\right.	
	\end{equation}
	where $\alpha_{2}$ and $\alpha_{3}$ are scalars. 
	
	\section{Results}
	In this section, we measure the performances of our method using a database of \textit{melanoma skin cancer} patients which is an open access database \cite{cin-Lezoray-2014-1}.
	It is recalled that our method makes it possible to estimate the distributions of oxyhemoglobin and deoxyhemoglobin separately in addition to those of melanin and shading contrary to all existing methods (as mentioned in the introduction). So, in absolute terms we cannot compare our performance with any of these methods. However, since most of these methods estimate melanin and hemoglobin (which is a mixture of oxyhemoglobin and deoxyhemoglobin), we can limit ourselves in the comparison to these two chromophores. For this, we opted for a comparison with the methods proposed in \cite{gong2012hemoglobin,inbook} because they are accessible for testing, unlike most of the existing methods \cite{liu2013melanin,spigulis2015snapshot,spigulis2017smartphone}. 
	As for the performance measurement criteria, in addition to the classical visual criterion which is a subjective criterion, we propose in this paper a new numerical criterion. In fact, for dermatologists, melanoma is characterized by a high distribution of melanin, an average distribution of deoxyhemoglobin and a very low distribution of oxyhemoglobin compared to healthy skin \cite{phdthesis}. To support this subjective criterion, which is the most used in the literature \cite{gong2012hemoglobin,liu2013melanin,spigulis2015snapshot,spigulis2017smartphone}, we use a second numerical criterion, which allows us to check to what extent the independence hypothesis we have assumed is satisfied by the estimated chromophores, since this hypothesis is the most used by researchers \cite{inbook,ojima2004application, mitra2010source}. We then define our new performance measurement criterion, which we note as $C_{Ind}$, by exploiting the assumption of statistical independence at order 4 between melanin and hemoglobin, as follows:
	\begin{equation}\label{eq4}
		C_{Ind}=\frac{1}{2}\left( C_{13}+C_{31}\right), 
	\end{equation}
	where:
	\begin{equation}%\label{eq4}
		C_{13}=20\,log_{10}\left( \left| E[\tilde{Y}_{1}(u)\tilde{Z}_{1}(u)^3]\right|^{-1} \right) 
		\hspace{0.2cm} \text{and}\hspace{0.2cm}
		C_{31}=20\,log_{10}\left( \left| E[\tilde{Y}_{1}(u)^3\tilde{Z}_{1}(u)]\right| ^{-1}\right)
	\end{equation}
	We recall that the estimates $\tilde{Y}_{1}(u)$ and $\tilde{Z}_{1}(u)$ are respectively the centered versions of the melanin and hemoglobin distributions.
	
\begin{figure}[h]
		\centering
		\includegraphics[width=3.5cm,height=2.3cm]{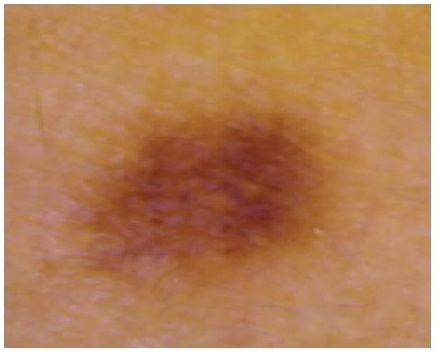}	
		\caption{Treated multispectral dermatological image.}
		\label{fig6}
	\end{figure}

	The database contains 30 multispectral images \cite{cin-Lezoray-2014-1}. Knowing that we have tested our method on some images and that the results obtained are similar, we limit ourselves here to present the results of a single melanoma image, due to lack of space. This image is shown in Figure \ref{fig6}.
		\vspace*{-0.5cm}
	\vspace*{-0.4cm}
	
\begin{figure}[h!]
		\centering
		\begin{tabular}{cccc}
			\vspace*{0.2cm}
			\includegraphics[width=0.2\linewidth]{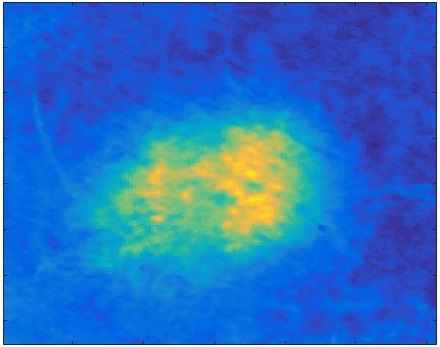}	&  	\includegraphics[width=0.2\linewidth]{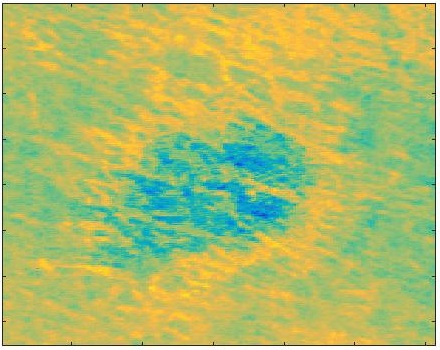}&  	\includegraphics[width=0.2\linewidth]{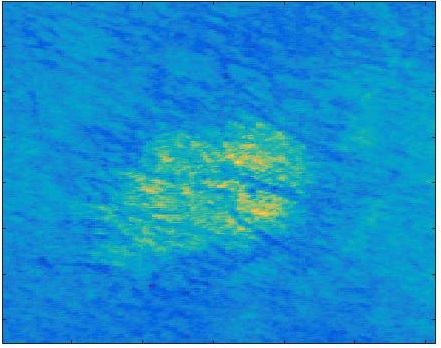}& 	\includegraphics[width=0.2\linewidth]{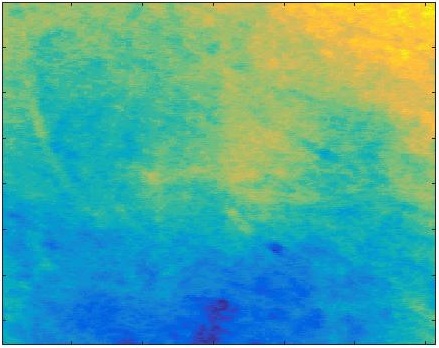} \\ 
			\includegraphics[width=0.25\linewidth]{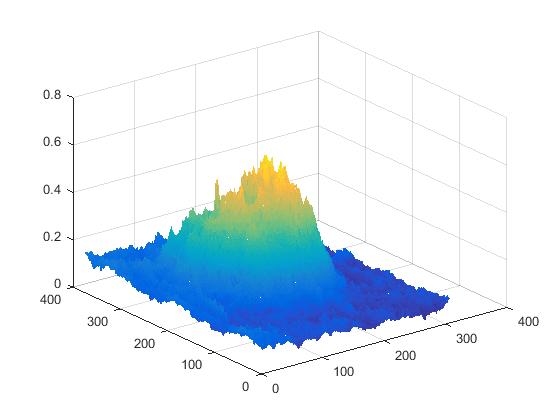}	&  	\includegraphics[width=0.25\linewidth]{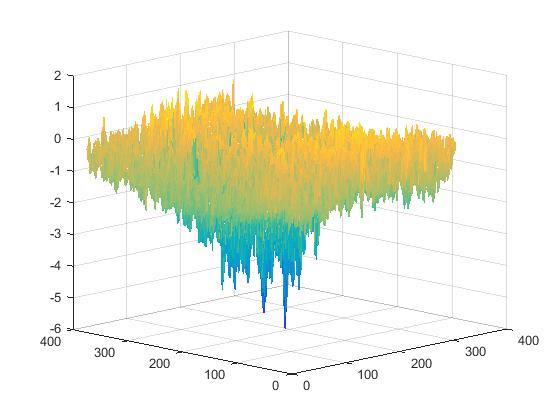}&  	\includegraphics[width=0.25\linewidth]{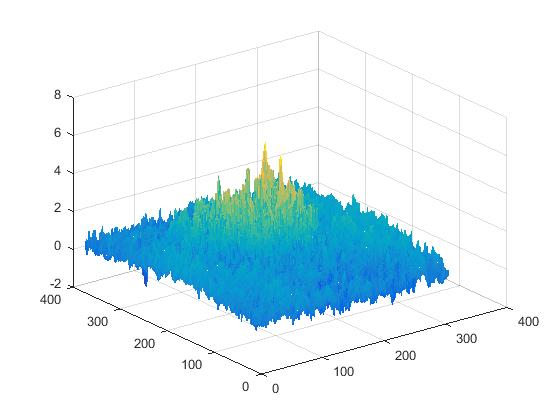}& 	\includegraphics[width=0.25\linewidth]{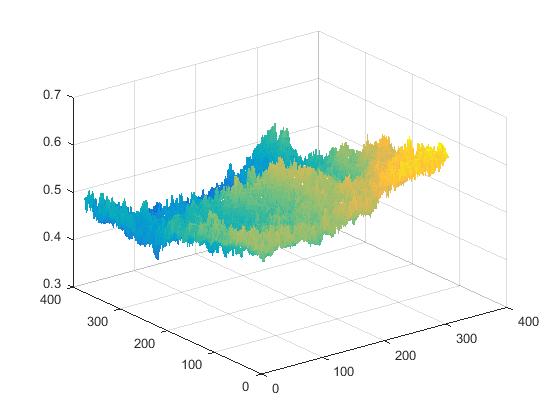} \\ 
			& \tcb{Oxyhemoglobin}&&\tcb{Shading}\\
			\tcb{Melanin} & &\tcb{Deoxyhemoglobin}&   \\  
		\end{tabular} 
		\caption{Estimated distributions (2D and 3D) of the three chromophores and shading using \textit{our method}.}\label{figOutPM}
	\end{figure}

\begin{figure}[h!]
		\centering
		\begin{tabular}{cc|cc}
			&&&\\
			\includegraphics[width=0.2\linewidth]{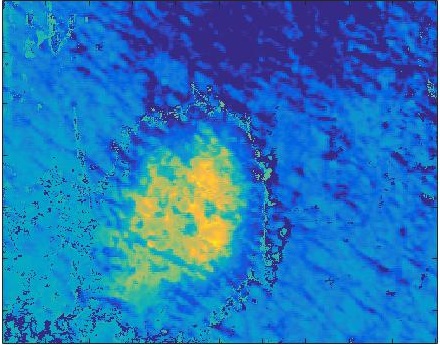}	& 	\includegraphics[width=0.2\linewidth]{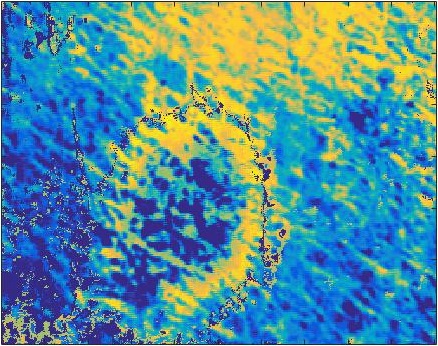}& 	\includegraphics[width=0.2\linewidth]{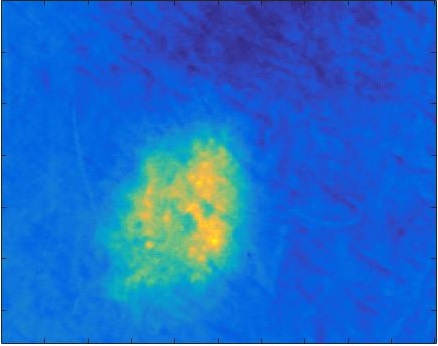}& 	\includegraphics[width=0.2\linewidth]{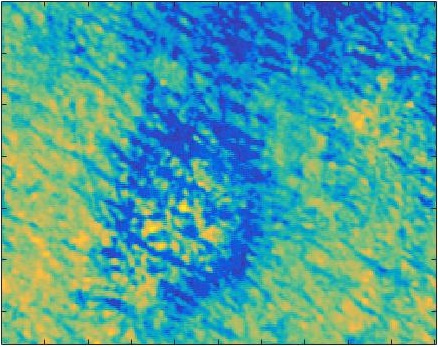} \\ 
			&&&\\
			\includegraphics[width=0.25\linewidth]{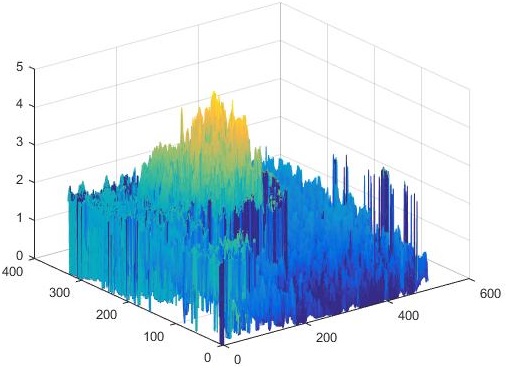}	&  	\includegraphics[width=0.25\linewidth]{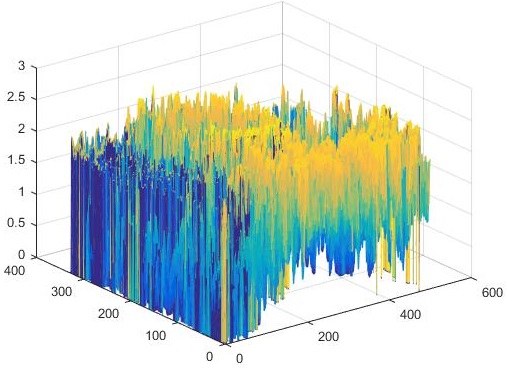}&  	\includegraphics[width=0.25\linewidth]{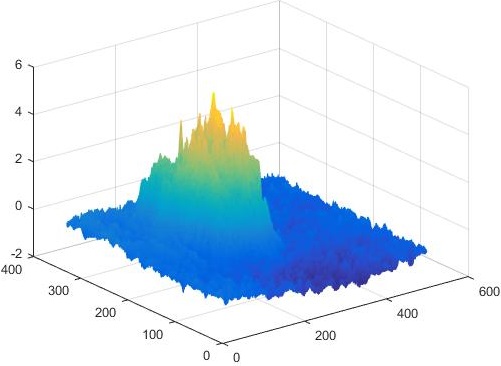}& 	\includegraphics[width=0.25\linewidth]{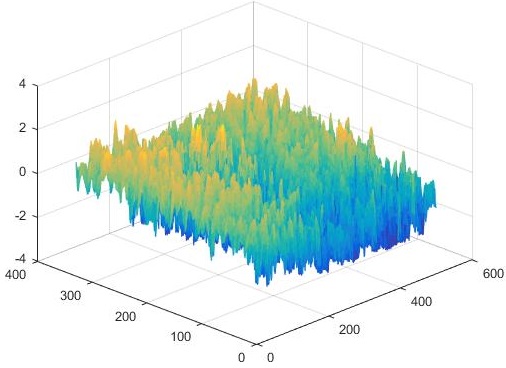} \\ 
			& \tcb{Hemoglobin}&&\tcb{Hemoglobin}\\
			\tcb{Melanin} & &\tcb{Melanin}&   \\  
		\end{tabular} 
		\vspace*{0.2cm}
		\begin{tabular}{cccc}
			(a) Method \cite{gong2012hemoglobin} \hspace*{1.74cm}	& & & \hspace*{1.7cm}(b) Method \cite{inbook}
		\end{tabular}
		\caption{Estimated distributions (2D and 3D) of melanin and hemoglobin using: (a) Method \cite{gong2012hemoglobin}, (b) Method \cite{inbook}.}\label{figOutPSM10}
	\end{figure}

	For the methods \cite{inbook,gong2012hemoglobin}, we used respectively the algorithm FastICA \cite{hyvarinen1997fast} and MU \cite{lee2000algorithms}. Two-dimensional (2D) and three-dimensional (3D) representations of the estimated distributions of each of the three chromophores in addition to the shading using our method are given in Figure \ref{figOutPM}. These 2D and 3D representations are grouped by column for each chromophore. The 2D and 3D representations of the estimated melanin and hemoglobin distributions using the methods \cite{gong2012hemoglobin,inbook} are grouped in Figure \ref{figOutPSM10}.
	
	Figure \ref{figOutPM} shows the good performance of our method for the estimation of the distribution of each of the three chromophores in addition to the shading. Indeed, we see that in the melanoma area we have a high distribution of melanin and a relatively high distribution of deoxyhemoglobin compared to oxyhemoglobin as shown in the 3D representations of these chromophores.
	This is fully consistent with the physiological knowledge that characterizes melanoma, since \textit{naevus} disease, which generally has a melanin distribution similar to that of \textit{melanoma}, is instead characterized by an oxyhemoglobin distribution that is close to that of deoxyhemoglobin \cite{phdthesis}.
	On the other hand, from Figure \ref{figOutPSM10}, we see that the melanin distribution estimated by the methods \cite{gong2012hemoglobin,inbook} is similar to that estimated by our method, which means that these two methods \cite{gong2012hemoglobin,inbook} also suspects melanoma.
	Nevertheless, its estimated hemoglobin distribution (which can be seen as a mixture of oxyhemoglobin, deoxyhemoglobin, and shading) does not allow to decide on the nature of the disease, since the distinction between melanoma and nevus can be made only by estimating the distributions of oxyhemoglobin and deoxyhemoglobin separately.
	All these findings, which were made on the basis of physiological knowledge of the melanoma disease, are in perfect agreement with the results obtained using our new numerical criterion, denoted $C_{Ind}$ and defined by the relation (\ref{eq4}), which are presented in the table \ref{Tab}. In this table we provide the mean and standard deviation of $C_{Ind}$, denoted $\overline{C}_{Ind}$ and $\sigma$, obtained on 10 different images from the database \cite{cin-Lezoray-2014-1}.
	\vspace*{-0.5cm}
	\begin{table}[H]
		\begin{center}
			\begin{tabular}{|c|c|c|c|}
				\hline
				&Method \cite{gong2012hemoglobin}&  Method \cite{inbook}      &   {Our method} \\
				\hline
				$\overline{C}_{Ind}\, (dB)$&   -2.05   &   27.45    &{\bf 38.44}\\
				\hline
				$\sigma\,(dB)$&  3.66   &     6.20    &{\bf 4.33}   \\
				\hline
			\end{tabular}
		\end{center}
		\caption{Mean and standard deviation of $C_{Ind}$ in $dB$.}
		\label{Tab}
	\end{table}
	\vspace*{-0.6cm}
	From the table \ref{Tab}, we see that our method is much better than the other methods since the value obtained for $\overline{C}_{Ind}$ using our method is larger than that obtained using the methods \cite{gong2012hemoglobin,inbook}. We also find that the \cite{gong2012hemoglobin} method has poor performance compared to our method and the method \cite{inbook} and this can be explained by the infinite solution problem posed by NMF. Since this criterion is based on a measure of independence, we deduce that our method provides estimates of melanin and hemoglobin distributions in output that are much more independent than those provided by the methods \cite{gong2012hemoglobin,inbook}.
	\section{Conclusion}
	In this paper we have proposed a new approach which aims to identify the melanoma diseases by identifying the distribution of its main skin chromophores (melanin, oxyhemoglobin and deoxyhemoglobin) from multispectral dermatological images. 
	The key idea of our approach is to take into account the shading, considering it as a full-fledged source, and the three chromophores, which leads to an instantaneous linear mixture model with four sources rather than two or three sources, as is the case for all existing methods. The results of all the tests carried out, using a database of real multispectral images of skin affected by a skin cancer of the type melanoma, have shown that our approach is very efficient using the classical criterion based on visual analysis than our new independence criterion. In terms of perspective, it would be interesting to test our method on other multispectral dermatological image databases of other skin diseases.

\end{document}